\documentclass[12pt]{article}

\usepackage{graphicx,psfrag,epsf}
\usepackage{enumerate}
\usepackage{natbib}
\usepackage{mathtools}
\usepackage{booktabs}
\usepackage{color}
\usepackage[english]{babel} 
\usepackage[protrusion=true,expansion=true]{microtype} 
\usepackage{amsmath,amsfonts,amsthm}
\usepackage{dsfont}
\usepackage{amssymb}
\usepackage{chngcntr}
\usepackage[toc,page]{appendix}
\usepackage{textcomp}
\usepackage{url}

\usepackage{array}
\usepackage{booktabs} 
\usepackage{natbib}
\usepackage{setspace}

\usepackage{soul}
\usepackage{multirow}
\usepackage{enumitem}
\usepackage{microtype} 
\usepackage[font = small,labelfont=bf,textfont=it]{subcaption}
\usepackage{xcolor}
\usepackage{tabularx}
\usepackage[font = small,labelfont=bf,textfont=it]{caption} 
\usepackage{footnote}
\usepackage{algorithm}
\usepackage[noend]{algpseudocode}
\usepackage{etoolbox}

\algblock{ParFor}{EndParFor}
\algnewcommand\algorithmicparfor{\textbf{for}}
\algnewcommand\algorithmicpardo{\textbf{do\ parallel}}
\algnewcommand\algorithmicendparfor{\textbf{end\ parallel\ for}}
\algrenewtext{ParFor}[1]{\algorithmicparfor\ #1\ \algorithmicpardo}
\algrenewtext{EndParFor}{\algorithmicendparfor}

\makeatletter
\def\BState{\State\hskip-\ALG@thistlm}

\newcommand{\distas}[1]{\mathbin{\overset{#1}{\kern\z@\sim}}}%
\newcommand{\bm}[1]{\mathbf{#1}}
\newcommand{\bb}[1]{\boldsymbol{#1}}
\newsavebox{\mybox}\newsavebox{\mysim}
\newcommand{\distras}[1]{%
  \savebox{\mybox}{\hbox{\kern3pt$\scriptstyle#1$\kern3pt}}%
  \savebox{\mysim}{\hbox{$\sim$}}%
  \mathbin{\overset{#1}{\kern\z@\resizebox{\wd\mybox}{\ht\mysim}{$\sim$}}}%
}
\newtheorem{theorem}{Theorem}

\newcommand{\be}{\begin{equation}}
\newcommand{\ee}{\end{equation}}
\newcommand{\bi}{\begin{itemize}}
\newcommand{\ei}{\end{itemize}}
\newcommand{\ben}{\begin{enumerate}}
\newcommand{\een}{\end{enumerate}}

\newcommand{\R}{\mathbb{R}}

\DeclarePairedDelimiter{\norm}{\lVert}{\rVert}

\newcolumntype{K}[1]{\geq {\centering\arraybackslash}p{#1}}
\allowdisplaybreaks
\DeclareMathOperator*{\argmin}{\arg\!\min}
\makeatother

\let\oldbibliography\thebibliography
\renewcommand{\thebibliography}[1]{\oldbibliography{#1}
\setlength{\itemsep}{0pt}} 

\newcommand{\blind}{0}

\patchcmd{\footnotemark}{\stepcounter{footnote}}{\refstepcounter{footnote}}{}{}

\usepackage{booktabs}

\usepackage{makecell}
\usepackage[section]{placeins}

\begin{document}

\def\spacingset#1{\renewcommand{\baselinestretch}%
{#1}\small\normalsize} \spacingset{1}

\if1\blind
{
  \title{\bf }
  \small
  \author{}\hspace{.2cm}\\
  \maketitle
} \fi

\if0\blind
{
  \bigskip
  \bigskip
  \bigskip
  \begin{center}
    {\LARGE\bf  Rational Kriging}\vspace{.2cm}\\
    {V. Roshan Joseph}\vspace{.2cm}\\
    {H. Milton Stewart School of Industrial and Systems Engineering\\ 
    Georgia Institute of Technology, Atlanta, GA 30332, USA\\
    roshan@gatech.edu}\vspace{.2cm}\\
\end{center}
  \medskip
} \fi
\vspace{.25in}
\bigskip

\vspace{-0.5cm}
\begin{abstract}
\noindent This article proposes a new kriging that has a rational form. It is shown that the generalized least squares estimate of the mean from rational kriging is much more well behaved than that from ordinary kriging. Parameter estimation and uncertainty quantification for rational kriging are proposed using a Gaussian process framework. Its potential applications in emulation and calibration of computer models are also discussed.
\end{abstract}


\spacingset{1.45} 

\section{Introduction} \label{sec:intro}
Kriging is a technique for multivariate interpolation of arbitrarily scattered data. It is originated from some mining-related applications, which is developed into the field of geostatistics by the pioneering work of \cite{matheron1963principles}. It has now become a prominent technique  for function approximation and uncertainty quantification in spatial statistics \citep{cressie2015statistics}, computer experiments \citep{Santner2003}, and machine learning \citep{rasmussen2006gaussian}.

Kriging can be briefly explained as follows. Suppose we have observed the data $\{(\bm x_i,y_i)\}_{i-1}^n$, where $\bm x\in \R^p$ is the $p$-dimensional inputs and $y$ the output. The aim is to predict $y$ for a future $\bm x$. To do this, construct a linear predictor $\hat{y}(\bm x)=\bm a(\bm x)'\bm y=\sum_{i=1}^n a_i(\bm x)y_i$. Kriging gives the best linear unbiased predictor (BLUP) under some assumptions of the data generating process. Specifically, if the data are generated from a second-order stationary stochastic process with mean $\mu$, variance $\tau^2$, and correlation function $cor\{Y(\bm u),Y(\bm v)\}=R(\bm u-\bm v)$, then the kriging predictor can be obtained by minimizing the mean squared prediction error \citep{Santner2003}
\[E\left\{Y(\bm x)-\hat{y}(\bm x)\right\}^2\]
with respect to $\bm a(\bm x)$ subject to the condition that $E\{\hat{y}(\bm x)\}=\mu$ for all $\bm x$. The optimal solution is given by 
\[\bm a(\bm x)'=\{1-\bm r(\bm x)'\bm R^{-1}\bm 1\} \frac{\bm 1'\bm R^{-1}}{\bm 1'\bm R^{-1}\bm 1} +\bm r(\bm x)'\bm R^{-1},\] 
where $\bm R=\{R(\bm x_i-\bm x_j)\}_{n\times n}$ is the correlation matrix, $\bm r(\bm x)=(R(\bm x-\bm x_1),\ldots,R(\bm x-\bm x_n))'$, and $\bm 1$ is a vector of $n$ 1's. Substituting the solution in the linear predictor and simplifying, we obtain the (ordinary) kriging predictor as
\begin{equation}\label{eq:ok}
    \hat{y}_{OK}(\bm x)=\hat{\mu}_{OK}+\bm r(\bm x)'\bm R^{-1}(\bm y-\hat{\mu}_{OK}\bm 1),
\end{equation}
where 
\begin{equation}\label{eq:muOK}
    \hat{\mu}_{OK}=\frac{\bm 1'\bm R^{-1}\bm y}{\bm 1'\bm R^{-1}\bm 1}.
\end{equation}
The expression in (\ref{eq:muOK}) shows that  $\hat{\mu}_{OK}$ is the well-known Generalized Least Squares (GLS) estimate of $\mu$.

\cite{joseph2006limit} noticed that the ordinary kriging predictor has sometimes a ``mean reversion'' issue and proposed a modified predictor
\begin{equation}\label{eq:lk}
     \hat{y}(\bm x)=\frac{\bm r(\bm x)'\bm R^{-1}\bm y}{\bm r(\bm x)'\bm R^{-1}\bm 1},
\end{equation}
whose predictions tend towards the nearest neighbor value when the correlations go to zero and thus, avoids the mean reversion issue. This predictor can be viewed as a limiting case of a simple kriging predictor with a recursive estimation of $\mu$ and hence it is called limit kriging. Different from ordinary kriging, limit kriging has a rational form. The purpose of this article is to examine optimal rational predictors of the form 
\begin{equation}\label{eq:rational}
    \hat{y}(\bm x)=\frac{\bm a(\bm x)'\bm y}{\bm a(\bm x)'\bm 1}.
\end{equation}

Although rational polynomials have a long history in  interpolation, its extension to radial basis functions (RBFs) is very recent. \cite{jakobsson2009rational} proposed to use rational RBFs for modeling resonance phenomena. \cite{sarra2018rational} also found that rational RBFs perform exceptionally well for modeling functions with discontinuities and steep gradients. In a more recent work, \cite{buhmann2020analysis} showed that rational RBFs have comparable approximation accuracy to the classical RBFs, but has more robust prediction performance. However, RBFs cannot provide any uncertainty quantification. In contrast, owing to its probabilistic formulation, kriging can automatically provide prediction intervals and can easily be integrated into Bayesian methods and non-normal data settings. 

\begin{figure}[h!]
\begin{center}
\begin{tabular}{ccc}
\includegraphics[width = .32\textwidth]{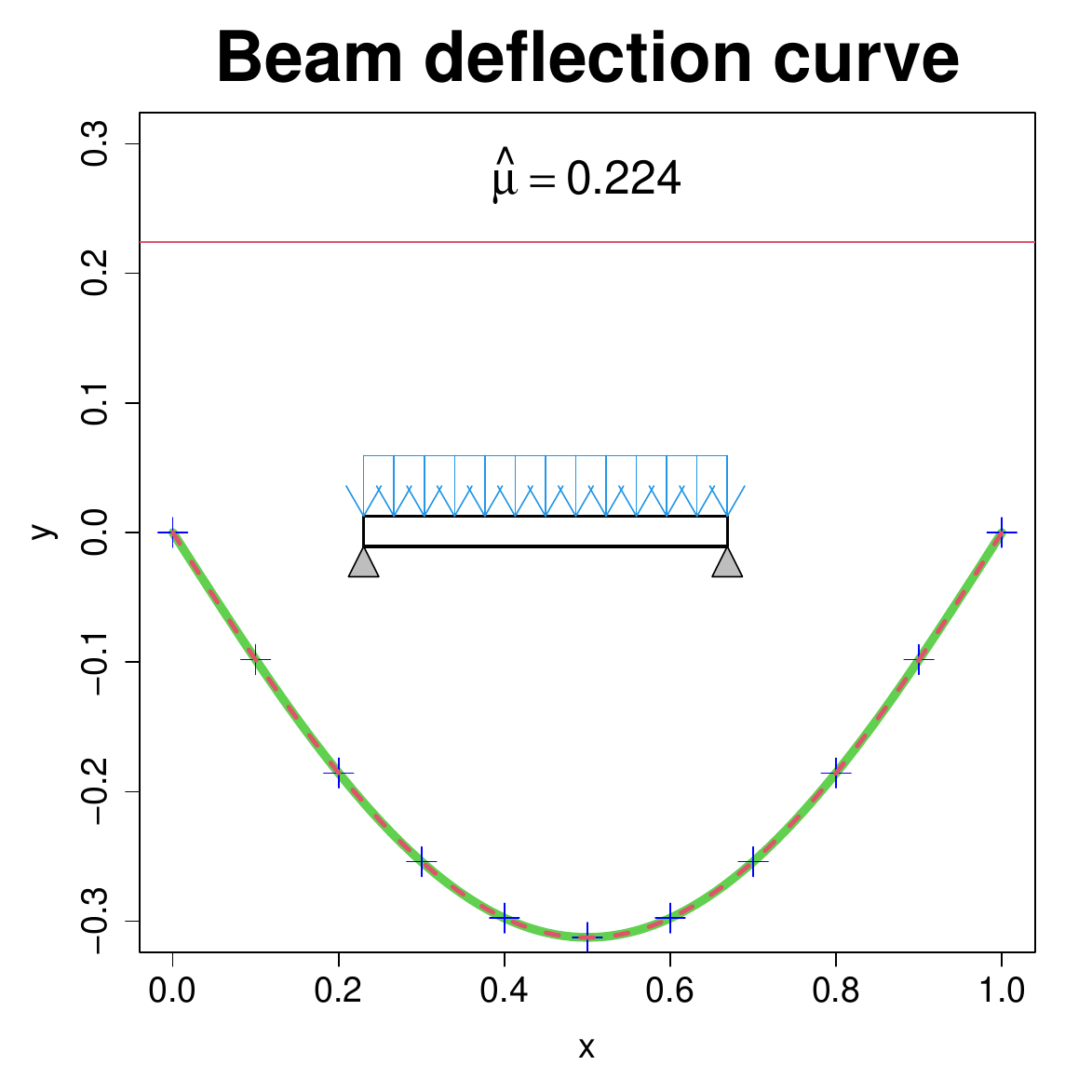}&
\includegraphics[width = .32\textwidth]{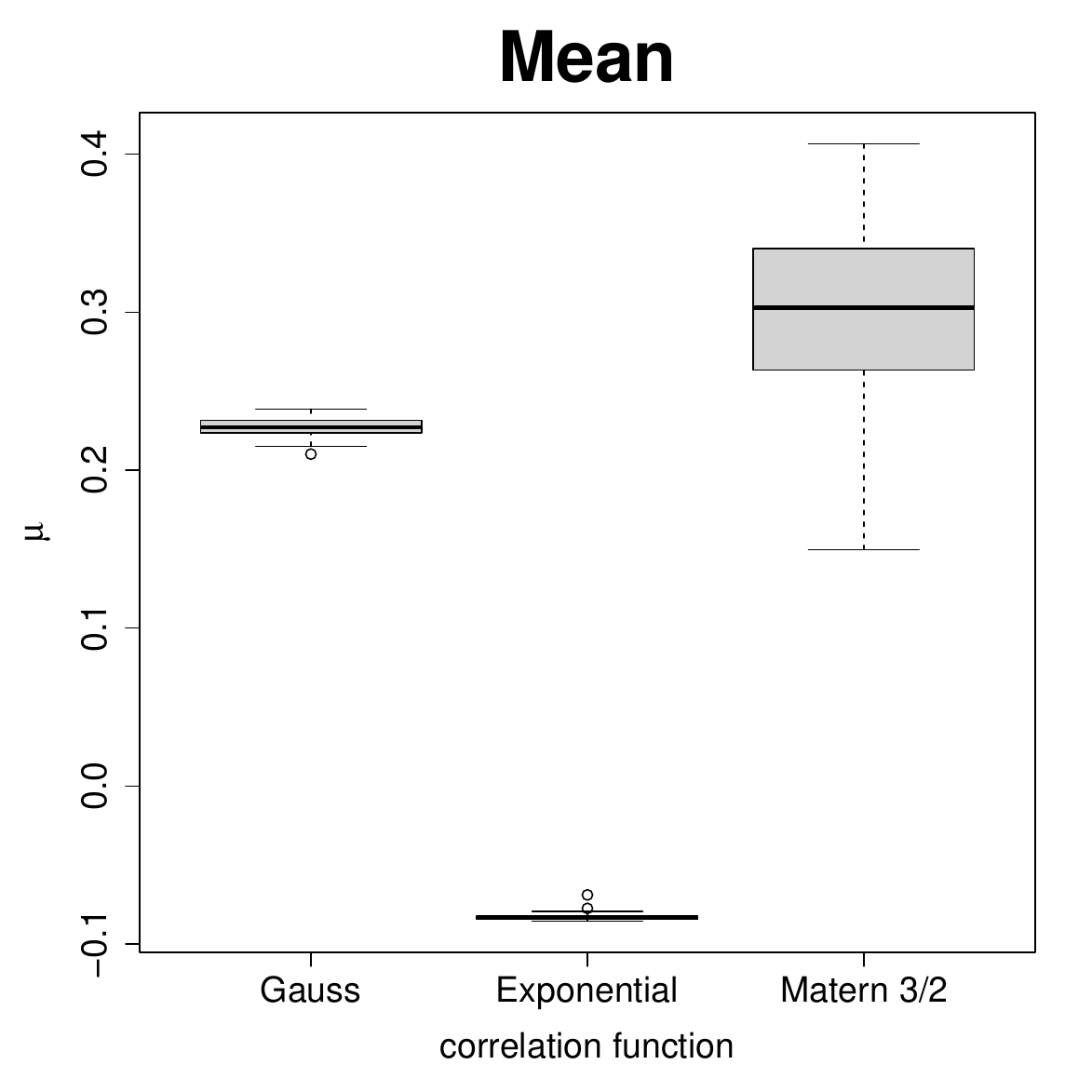}& 
\includegraphics[width = .32\textwidth]{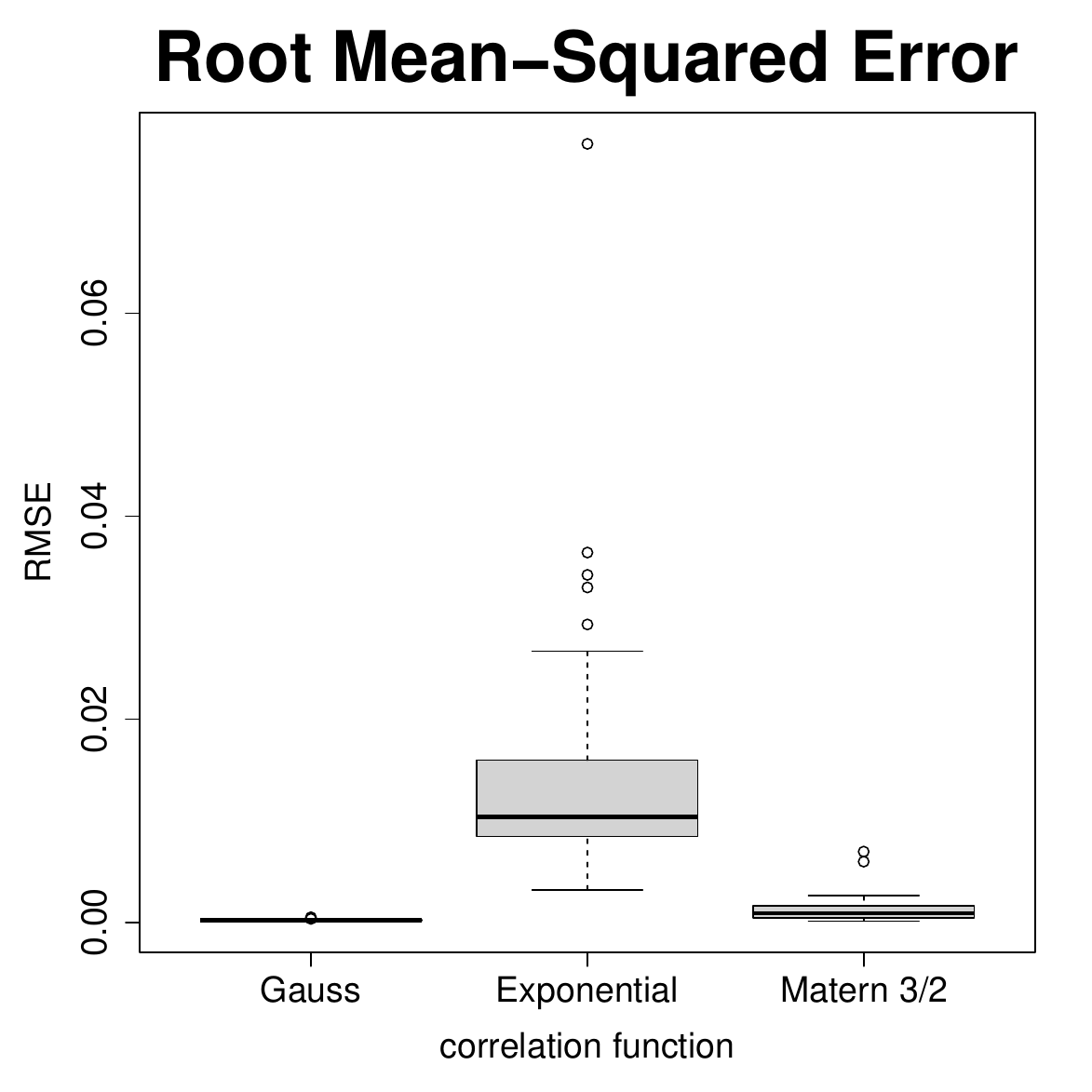} 
\end{tabular}
\caption{(left) Plot of the beam deflection curve (solid-green), data (blue-pluses), and ordinary kriging predictor (dashed-red). A simply supported beam with uniform load is shown as an inset of this plot. (middle) Boxplot of $\hat{\mu}$'s from 50 simulations using three correlation functions where the $x_i$'s are randomly sampled in $[0,1]$. (right) Root mean squared errors from the 50 simulations.}
\label{fig:example1}
\end{center}
\end{figure}

Different from the RBF literature, we will motivate the benefit of rational predictors using parameter estimation accuracy. As an example, consider the deflection of a simply supported beam with uniform load shown as an inset in the left panel of Figure \ref{fig:example1}. The deflection at a distance $x$ from the left end of the beam is given by 
\[y=-\frac{P}{24EI}x(x^3-2Lx^2+L^3),\]
where $P$ is the uniform load density, $E$ is the elastic modulus, $I$ is the area moment of inertia, $L$ is the length of the beam, and $x\in [0,L]$. Let $P/(24EI)=1$ and $L=1$. The function is plotted in the left panel of Figure \ref{fig:example1} along with 11 equi-spaced $x_i$'s from $[0,1]$. An ordinary kriging was fitted to this data with a Gaussian correlation function $R(h)=e^{- (h/\theta)^2}$, where the unknown correlation parameter $\theta$ is estimated from the data using maximum likelihood. We used the R package \texttt{DiceKriging} \citep{roustant2012dicekriging} for estimation, where a small nugget of $10^{-6}$ is applied for numerical stability. The predictions in $[0,1]$ are plotted in the left panel of Figure \ref{fig:example1}, which are almost indistinguishable with the true function values showing excellent prediction performance. The GLS estimate of $\mu$ from (\ref{eq:muOK}) is obtained as $\hat{\mu}_{OK}=0.224$. Interestingly, this value is outside the range of the observed function values, which are from $[-0.3125,0]$.

We repeated this exercise 50 times by uniformly sampling $x_i$'s from $[0,1]$ and using two more correlation functions: exponential and Mat\'ern 3/2 \citep[p.84]{rasmussen2006gaussian}. We can see from the middle panel of Figure \ref{fig:example1} that the estimates of $\mu$ using exponential correlation function are around $-0.1$, which are in the range of the observed values, but the estimates of $\mu$ from the Mat\'ern 3/2 correlation function are generally higher than those obtained from the Gaussian correlation function. For each simulation, the root mean squared error (RMSE) is calculated over a grid of 1,001 values and is shown in the right panel of Figure \ref{fig:example1}. We can see that the Gaussian correlation function gives the best prediction in this example. The prediction from the exponential correlation function is the worst in spite of having the mean in the observed range of $y_i$'s.

Although better prediction is obtained when $\hat{\mu}$ is outside the observed  range of function values,  the interpretation of those estimates becomes questionable. One could argue that $\mu$ is the mean of a stochastic process in which the beam deflection curve is just a realization and thus a value around $0.2$ is an admissible estimate. However, if $\mu$ has a physical interpretation, then this estimate is meaningless. For example, a positive value of mean would imply that the beam will deflect in the opposite direction of the force, which is against the law of nature! This is a common dilemma in model calibration problems \citep{kennedy2001bayesian}. We will show that the use of rational kriging can surprisingly avoid this issue without sacrificing the prediction performance.

A quick fix to the foregoing issue is to estimate the mean using ordinary least squares \citep{pronzato2023blue}. However, it leads to inconsistencies in the modeling framework-- an uncorrelated process for estimation and a correlated process for prediction. \cite{plumlee2018orthogonal} argued that the estimation problems are caused by identifiability issues between the stochastic process and the mean function (a constant function in the case of ordinary kriging). They proposed to overcome the identifiability issue by making the stochastic process orthogonal to the mean function. Although their approach is very general, it leads to a nonstationary correlation function that involves high-dimensional integrals making the estimation computationally challenging and numerically unstable. In contrast, rational kriging requires only a rescaling of the original predictor, which is very easy to implement in practice. 

The article is organized as follows. Section 2 develops the rational kriging and a Gaussian process framework for parameter estimation and uncertainty quantification. Simulations with several test functions are provided in Section 3. Potential applications of rational kriging in emulation and calibration of computer models are illustrated with some examples in Section 4. Some concluding remarks are given in Section 5.

\section{Methodology}
We will first derive the optimal rational predictor and then investigate its estimation properties by assuming Gaussianity for the stochastic process.

\subsection{Rational Kriging}
Let $\mathcal{X}$ be the input region for data collection. Most of the time, it can be scaled in $[0,1]^p$. Notice that this region does not come into the formulation or derivation of the ordinary kriging predictor because we assume the stationary stochastic process has a constant mean $\mu$ and variance $\tau^2$ for all $\bm x\in \R^p$. This could be the reason why the estimate of $\mu$ went outside the observed range of $y$ values in the example that we saw earlier. So, we can possibly overcome the issue by assuming a nonstationary variance $\tau^2(\bm x)$, where it should increase  as $\bm x$ goes outside of $\mathcal{X}$. 

Now consider a rational predictor of the form
\begin{equation}\label{eq:rational2}
    \hat{y}(\bm x)=\frac{\bm a(\bm x)'\bm y}{b(\bm x)},
\end{equation}
where $\bm a(\bm x)=(a_1(\bm x),\ldots,a_n(\bm x))'$ and $b(\bm x)$ are functions of the input variables $\bm x=(x_1,\ldots,x_p)'$. Assume that the data $\bm y$ is a realization from a second-order stationary stochastic process with mean $\mu$, variance $\tau^2(\bm x)$, and correlation function $R(\cdot)$. Then, the  predictor in (\ref{eq:rational2}) will be unbiased if
\[E\{\hat{y}(\bm x)\}=\frac{\bm a(\bm x)'E\{\bm y\}}{b(\bm x)}=\mu\frac{\bm a(\bm x)'\bm 1}{b(\bm x)}=\mu\]
for all $\bm x$, which implies $b(\bm x)=\bm a(\bm x)'\bm 1$. Now we can find the best rational unbiased predictor by minimizing
\[MSPE=E\left\{Y(\bm x)- \frac{\bm a(\bm x)'\bm y}{\bm a(\bm x)'\bm 1}\right\}^2\]
with respect to $\bm a(\bm x)$. It is easy to show that $cov(Y(\bm x),\bm y)=\tau(\bm x) diag(\bb \tau)\bm r(\bm x)$ and $var\{\bm y\}=diag(\bb \tau)\bm R diag(\bb \tau)$, where $\bb \tau=(\tau(\bm x_1),\ldots,\tau(\bm x_n))'$ and $diag(\bb \tau)$ is a diagonal matrix with diagonal elements $\bb \tau$. Thus,
\[MSPE=\tau^2(\bm x)-2\frac{\bm a(\bm x)'}{\bm a(\bm x)'\bm 1}\tau(\bm x)diag(\bb \tau)\bm r(\bm x)+\frac{\bm a(\bm x)'}{\bm a(\bm x)'\bm 1}diag(\bb \tau)\bm R diag(\bb \tau)\frac{\bm a(\bm x)}{\bm a(\bm x)'\bm 1}.\]
Differentiating with respect to $\bm a(\bm x)$ and equating to zero, we obtain
\[diag(\bb \tau)\left\{ -2\tau(\bm x) \bm r(\bm x)+2\bm R diag(\bb \tau)\frac{\bm a(\bm x)}{\bm a(\bm x)'\bm 1}\right\}\frac{\partial}{\partial \bm a(\bm x)}\left(\frac{\bm a(\bm x)}{\bm a(\bm x)'\bm 1}\right) =0.\]
Thus,
\[\frac{\bm a(\bm x)}{\bm a(\bm x)'\bm 1}= \tau(\bm x)diag(\bb \tau^{-1})\bm R^{-1}\bm r(\bm x)\]
is a solution, provided $\tau(\bm x) \bm r(\bm x)'\bm R^{-1}diag(\bb \tau^{-1})\bm 1=1$. Therefore, we can let
\begin{equation}\label{eq:tau}
    \tau(\bm x)=\frac{1}{\bm r(\bm x)'\bm R^{-1}(\bm 1/\bb \tau)},
\end{equation}
where $\bm 1/\bb\tau=(1/\tau_1,\ldots,1/\tau_n)'$ and $\tau_i=\tau(\bm x_i)$ for $i=1,\ldots,n$. Since $\bm r(\bm x_i)'\bm R^{-1}=(0,\ldots,1,\ldots,0)'$ with $1$ at the $i$th position, (\ref{eq:tau}) holds for $i=1,\ldots,n$. However, (\ref{eq:tau}) is meaningful only if $\bm r(\bm x)'\bm R^{-1}(\bm 1/\bb \tau)>0$ for all $\bm x$. This can be ensured by putting a constraint on $\bb \tau\ge 0$ such that $\bm R^{-1}(\bm 1/\bb \tau)\geq 0$ and choosing a correlation function that does not vanish. Interestingly, $\tau(\bm x)\rightarrow \infty$ as $||\bm x-\bm x_i||\rightarrow \infty$ for all $i$, which agrees with our intuition.

Thus, we obtain the optimal rational kriging predictor as
\begin{equation} \label{eq:rk}
    \hat{y}(\bm x)=\frac{\bm r(\bm x)'\bm R^{-1}(\bm y/\bb \tau)}{\bm r(\bm x)'\bm R^{-1}(\bm 1/\bb \tau)}.
\end{equation}
We can see that the limit kriging predictor in (\ref{eq:lk}) is a special case of this predictor with $\bb \tau=\bm 1$. However, limit kriging is not an admissible predictor in the new formulation because $\bm R^{-1}\bm 1$ is not guaranteed to be nonnegative. The new predictor has $n$ additional unknown parameters $\bb \tau=(\tau_1,\ldots,\tau_n)'$, which can be chosen to ensure that $\bm R^{-1}(\bm 1/\bb \tau)\geq 0$ and $\bb \tau\ge 0$. 

Let $\bm R^{-1}(\bm 1/\bb \tau)=\bm c/\nu$, where $\bm c\ge 0$ and $\nu$ is a positive constant. Since $\bm R$ is a positive matrix, $\bm c\ge0$ implies $\bm 1/\bb \tau=\bm R\bm c/\nu\ge 0$. Thus, the rational kriging predictor can be written as
\begin{equation}\label{eq:rk2}
    \hat{y}(\bm x)=\frac{\bm r(\bm x)'\bm R^{-1}diag(\bm R\bm c)\bm y}{\bm r(\bm x)'\bm c},
\end{equation}
where $\bm c\ge0$. This is the same predictor obtained by \cite{kang2016kernel} as the limiting case of an iterated kernel regression. The choice of $\bm c$ will be discussed in the next section.

The derivation of rational kriging predictor does not give any estimate of $\mu$. However, since $\bm y$ is a random vector with mean $\mu\bm 1$ and variance $diag(\bb \tau)\bm R diag(\bb \tau)$, we can use the GLS estimate for $\mu$:
\begin{eqnarray}
\hat{\mu} &=& \frac{\bm 1'diag(\bb \tau^{-1})\bm R^{-1} diag(\bb \tau^{-1})\bm y }{\bm 1'diag(\bb \tau^{-1})\bm R^{-1} diag(\bb \tau^{-1})\bm 1}\nonumber\\
&=&\frac{\bm c'diag(\bm R\bm c)  \bm y}{\bm c'\bm R \bm c}.\label{eq:muhat}
\end{eqnarray}
Since $\bm c\ge 0$, we have the following result, which is in stark contrast to the GLS estimate of $\mu$ in ordinary kriging, where it can go outside the range of the data as we have observed in an example in the introduction. 
\begin{theorem} 
In rational kriging, the GLS estimate of $\mu$ is a convex combination of $\{y_i\}_{i=1}^n$ and therefore, it will always be in the range $[\min_i y_i,\max_i y_i]$ for any positive definite correlation function.
\end{theorem}

In order to understand if the GLS estimate from rational kriging is good or not, we
need to define the notion of a ``true value'' for $\mu$. Define the true value as the $L_2$-projection of the underlying function as in \cite{tuo2015efficient}:
\[\mu^*=\argmin_{\mu} \int_{\mathcal{X}} \{y(\bm x)-\mu\}^2 dF(\bm x)= \int_{\mathcal{X}} y(\bm x) dF(\bm x),\]
where $F(\cdot)$ is the distribution function of $\bm x$ with support $\mathcal{X}$ from which the input points are generated. Since $\mu^*$ is a convex combination of the $y(\bm x)$ for all $\bm x\in \mathcal{X}$, we can expect the rational kriging estimate $\hat{\mu}$ to be closer to $\mu^*$ than  $\hat{\mu}_{OK}$ to $\mu^*$. We will investigate this more in Section 3 using simulations.

The mean squared prediction error for the optimal rational kriging predictor is given by
\[MSPE=\frac{\nu^2}{\{\bm r(\bm x)'\bm c\}^2}\{1-\bm r(\bm x)'\bm R^{-1}\bm r(\bm x)\},\]
which can be used for uncertainty quantification. It can be computed only after specifying the parameter $\nu$ and the coefficients $\bm c$. Moreover, there are unknown parameters in the correlation function that need to be specified. We will develop their estimation procedure after introducing Gaussian Process (GP) in the next section.

\subsection{Rational Gaussian Process}
It is well known that the ordinary kriging predictor can be obtained as the posterior mean if we assume a GP prior for the true function that generated the data \citep{currin1991bayesian, rasmussen2006gaussian}. A similar framework can be developed for rational kriging. Following \cite{kang2016kernel}, assume
\begin{equation}
    y(\bm x)=\mu+\frac{\nu}{\bm r(\bm x)'\bm c} Z(\bm x), \;\; Z(\bm x)\sim GP(0,R(\cdot)).
\end{equation}
It is easy to show that
\begin{equation}\label{eq:post}
    y(\bm x)|\bm y\sim N\left(\hat{y}(\bm x), \frac{\nu^2}{\{\bm r(\bm x)'\bm c\}^2}\{1-\bm r(\bm x)'\bm R^{-1}\bm r(\bm x)\} \right),
\end{equation}
where $\hat{y}(\bm x)$ is the rational kriging predictor given in (\ref{eq:rk2}). As alluded to in the introduction, (\ref{eq:post}) can be used for constructing the prediction intervals, which is a major advantage of GPs over RBFs.

There are several unknown parameters in (\ref{eq:post}): $\mu$, $\nu$, and $\bm c$. In addition, the correlation functions have unknown parameters; denote them by $\bb \theta$. Among all these parameters, we will give a fully Bayesian treatment only for $\mu$. All the other parameters will be estimated or specified as follows.

The likelihood is given by
\[\bm y|\mu, \nu, \bm c, \bb \theta \sim N(\mu \bm 1,\nu^2 diag(\bm 1/\bm R\bm c)\;\bm R \;diag(\bm 1/\bm R\bm c)).\]
Assuming a non-informative prior for $\mu$: $p(\mu)\propto 1$, we obtain
\[\mu|\bm y, \nu, \bm c, \bb \theta \sim N\left( \hat{\mu}, \frac{\nu^2}{\bm c'\bm R\bm c }\right),\]
where $\hat{\mu}$ is the GLS estimate of $\mu$ given in (\ref{eq:muhat}). Looking at the posterior variance of $\mu$, it is tempting to choose $\bm c$ to maximize $\bm c'\bm R\bm c$. In fact, an elegant solution to this optimization problem exists. Under the constraint $\norm{\bm c}_2=1$, the quadratic form $\bm c'\bm R\bm c$ is maximized by the eigenvector corresponding to the largest eigenvalue of $\bm R$. Since $\bm R$ is a positive matrix,  this eigenvector is positive by Perron's theorem \citep{perron1907theorie}. Thus, we have the following result.

\vspace{.1in}
\noindent {\bf Proposition 1.} \emph{The posterior variance of $\mu$ can be minimized by taking $\bm c$ to be the eigenvector of $\bm R$ corresponding to its largest eigenvalue.}
\vspace{.1in}

\cite{buhmann2020analysis} also suggests to use this estimate for $\bm c$. Their suggestion is based on minimizing the native space norm of functions with kernel $K(\bm u,\bm v)=R(\bm u-\bm v)$. With this choice of $\bm c$, $\bm r(\bm x)'\bm c$ can be viewed as the Nystr\"{o}m approximation of the first eigenfunction of $R(\cdot)$ \cite[Sec. 4.3.2]{rasmussen2006gaussian}. In our trials, we found this estimate of $\bm c$ to work well when the functions are smooth, but poorly  when the functions are non-smooth. This is because $\bm r(\bm x)'\bm c$ can become very small for some value of $\bm x$, which can make the predictions erratic.

Another possibility is to let $\bm c=\bm R^{-1}\bm 1$ as in limit kriging \citep{joseph2006limit}, but this does not ensure nonnegativity of $\bm c$. We can overcome the nonnegativity issue as follows. Let $\hat{\gamma}$ be the smallest $\gamma \in [0,1]$ such that $[(1-\gamma)\bm R+\gamma \bm I]^{-1}\bm 1\ge \Delta \bm 1$ component-wise, where $\Delta\in [0,1]$.
Such a $\hat{\gamma}$ always exists because $\gamma=1$ trivially satisfies the inequality. Therefore, let
\begin{equation}\label{eq:chat}
    \hat{\bm c}=[(1-\hat{\gamma})\bm R+\hat{\gamma} \bm I]^{-1}\bm 1.
\end{equation}
Empirically, we found that $\Delta=\lambda_1/n$ works well, where $\lambda_1$ is the largest eigenvalue of $\bm R$.

When correlations are high, $\bm R\approx \lambda_1\bm E_1\bm E_1'$, where $\bm E_1$ the eigenvector corresponding to $\lambda_1$. Then, 
\[\bm R\hat{\bm c}\approx \frac{\lambda_1}{(1-\hat{\gamma})\lambda_1+\hat{\gamma}}\bm E_1\bm E_1'\bm 1\propto \lambda_1 \bm E_1=\bm R\bm E_1. \]
That is, the solution given in (\ref{eq:chat}) behaves exactly like the eigenvector solution of \cite{buhmann2020analysis}  when correlations are high (smooth functions). On the other hand, when correlations are small (nonsmooth functions), $\hat{\bm c}\approx [(1-\hat{\gamma})\bm I+\hat{\gamma} \bm I]^{-1}\bm 1 \propto \bm 1$, whereas $\bm E_1$ will be approximately the unit vector $(1,0,\ldots,0)'$. When this happens, the eigenvector solution will make  $\bm r(\bm x)'\bm c \approx 0$ for a large portion of $\mathcal{X}$, whereas $\bm r(\bm x)'\hat{\bm c} \approx 1$ for $\bm x$ in the neighborhood of the observed data points. Thus, the solution given in (\ref{eq:chat}) will be better behaved in all correlations regimes and therefore, will be adopted in this article. We also note that this solution is quite different from that of \cite{kang2016kernel}, where they estimated $\bm c$ by maximizing the unnormalized posterior, which is computationally prohibitive.

Thus, 
\begin{eqnarray*}
    p(\nu, \bb \theta|\bm y, \hat{\bm c})&\propto & \int p(\bm y|\mu,\nu,\hat{\bm c},\bb \theta)\;d\mu\\
    &\propto & \frac{|diag(\bm R\hat{\bm c})|}{\nu^{n-1}|\bm R|^{1/2}(\hat{\bm c}\bm R\hat{\bm c})^{1/2}}\exp{\left\{-\frac{1}{2\nu^2}(\bm y-\hat{\mu}\bm 1)'diag(\bm R\hat{\bm c})\bm R^{-1}diag(\bm R\hat{\bm c}) (\bm y-\hat{\mu}\bm 1)\right\}},
\end{eqnarray*}
where $\hat{\bm c}$ is given in (\ref{eq:chat}). Maximizing this with respect to $\nu$ and $\bb \theta$, we obtain
\begin{eqnarray}
    \hat{\nu}^2 &=& \frac{1}{n-1}(\bm y-\hat{\mu}\bm 1)'diag(\bm R\hat{\bm c})\bm R^{-1}diag(\bm R\hat{\bm c}) (\bm y-\hat{\mu}\bm 1),\\
    \hat{\bb \theta} &=&\argmin_{\bb \theta}\left\{ (n-1)\log\hat{\nu}^2+\log |\bm R|-2\sum_{i=1}^n \log (\bm R_i\hat{\bm c}) +\log (\hat{\bm c}'\bm R\hat{\bm c})\right\},
\end{eqnarray}
where $\bm R_i$ is the $i$th row of $\bm R$.

\subsection{Correlation Functions}
Rational kriging or rational GP can be used with any positive definite correlation function. One of the most commonly used correlation function in computer experiments is the Gaussian correlation function given by $R(\bm h) = \exp\{-\sum_{i=1}^p (h_i/\theta_i)^2\}$. Let $\theta_i^2=\theta^2/w_i$, where $\sum_{i=1}^pw_i=1$ and $w_i\ge 0$ for $i=1,\ldots,n$. Then the Gaussian correlation function can be written as
\begin{align}
    R(\bm h) &= \exp\left\{-\norm{\bm h}_w^2/\theta^2\right\},
\end{align}
where $\norm{\bm h}_w^2=\sum_{i=1}^p w_ih_i^2$. It is interesting to study the properties of the rational kriging predictor when the length-scale parameter ($\theta$) becomes small. Using a result in \cite{kang2016kernel}, it is easy to show that  the rational kriging tends to the nearest neighbor predictor defined by the norm $\norm{\cdot}_w$ as $\theta\rightarrow 0$. This property helps rational kriging to overcome the ``mean reversion'' problem commonly observed with ordinary kriging.

There is another correlation function that makes the foregoing limiting case even more interesting. Consider the rational quadratic function \citep{rasmussen2006gaussian} (also known as Cauchy function) given by
\begin{equation}\label{eq:rq}
    R(\bm h)=\left(1+\norm{\bm h}_{w}^2/\theta^2\right)^{-1}.
\end{equation}
When the length-scale parameter $\theta\rightarrow 0$, we have $\bm R\rightarrow \bm I$ and therefore $\hat{\bm c}\rightarrow \bm 1$. Moreover, $R(\bm x-\bm x_i)/\bm r(\bm x)'\hat{\bm c}\rightarrow \norm{\bm x-\bm x_i}_w^{-2}/\sum_{j=1}^n \norm{\bm x-\bm x_j}_w^{-2}$. The predictor
\[\hat{y}_{IDW}(\bm x)=\frac{\sum_{i=1}^n\norm{\bm x-\bm x_i}_w^{-2}y_i}{\sum_{j=1}^n \norm{\bm x-\bm x_j}_w^{-2}}\]
is the well-known inverse distance weighting (IDW) predictor \citep{shepard1968two, joseph2011regression}. Thus, we have the following result.

\begin{theorem} 
Under rational quadratic correlation function in (\ref{eq:rq}), the rational kriging predictor converges to the inverse distance weighting predictor as the length-scale parameter goes to 0.
\end{theorem}


\section{Simulations}

\subsection{One-dimensional function}
\begin{figure}[h]
\begin{center}
\includegraphics[width = 1\textwidth]{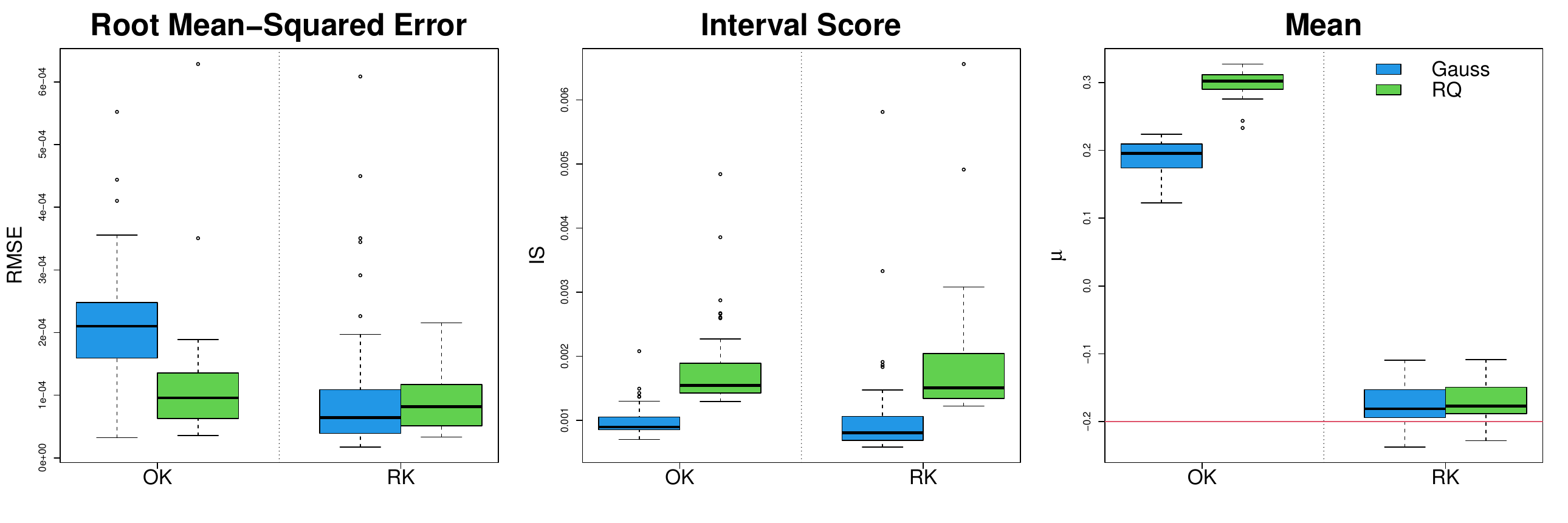} 
\caption{Boxplots of RMSE (left), IS (middle), and $\hat{\mu}$ (right) from the simulation using the beam deflection function. The simulation is done by randomly sampling $\{x_i\}_{i=1}^{11}$ from $[0,1]$. The true value $\mu^*$ is plotted as a red line in the right plot.}
\label{fig:beam}
\end{center}
\end{figure}

Consider again the beam deflection function used in the introduction: $y=-x(1-2x^2+x^3)$ for $x\in [0,1]$. Let $x_i\overset{iid}{\sim} U(0,1)$ for $i=1,\ldots,11$. These points are re-scaled such that $x_1=0$ and $x_{11}=1$. Ordinary kriging is fitted to the data using the Gaussian correlation function $R(h)=\exp\{-(h/\theta)^2\}$ and Rational Quadratic function $R(h)=\{1+(h/\theta)^2\}^{-1}$. Rational Kriging (RK) is also fitted to the same data using both the correlation functions following the procedure in Section 2.2. This simulation is repeated for 50 times. The left panel of Figure \ref{fig:beam} shows the Root Mean-Squared Errors (RMSEs) computed over a grid of 1,001 points in $[0,1]$. It shows that, on the average, rational kriging  is more accurate than ordinary kriging. Interval Score \citep{gneiting2007strictly} 
\[IS=\frac{1}{N}\sum_{i=1}^N\left[(u-l)+\frac{2}{\alpha}\{(l-t_i)_++(t_i-u)_+\}\right]\]
is computed for assessing the accuracy of $(1-\alpha)$ confidence intervals $[l,u]$, where $(x)_+=x$ if $x>0$ and 0 otherwise, and $\{t_i\}_{i=1}^N$ are the $N=1001$ testing locations. This is shown in the middle panel of Figure \ref{fig:beam} for 95\% confidence intervals. A small IS value indicates better confidence intervals (small width at prescribed coverage). In this example, IS shows comparable performance for rational kriging and ordinary kriging. The most striking result is the plot on the right panel of Figure \ref{fig:beam}. While ordinary kriging produces $\hat{\mu}$'s much larger than the maximum value of $y_i$'s, the estimate from rational kriging is around the true value  $\mu^*=\int_0^1 sin(2x)\; dx =0.708$ (shown as a red line in the same figure). 

\begin{figure}[h!]
\begin{center}
\includegraphics[width = .5\textwidth]{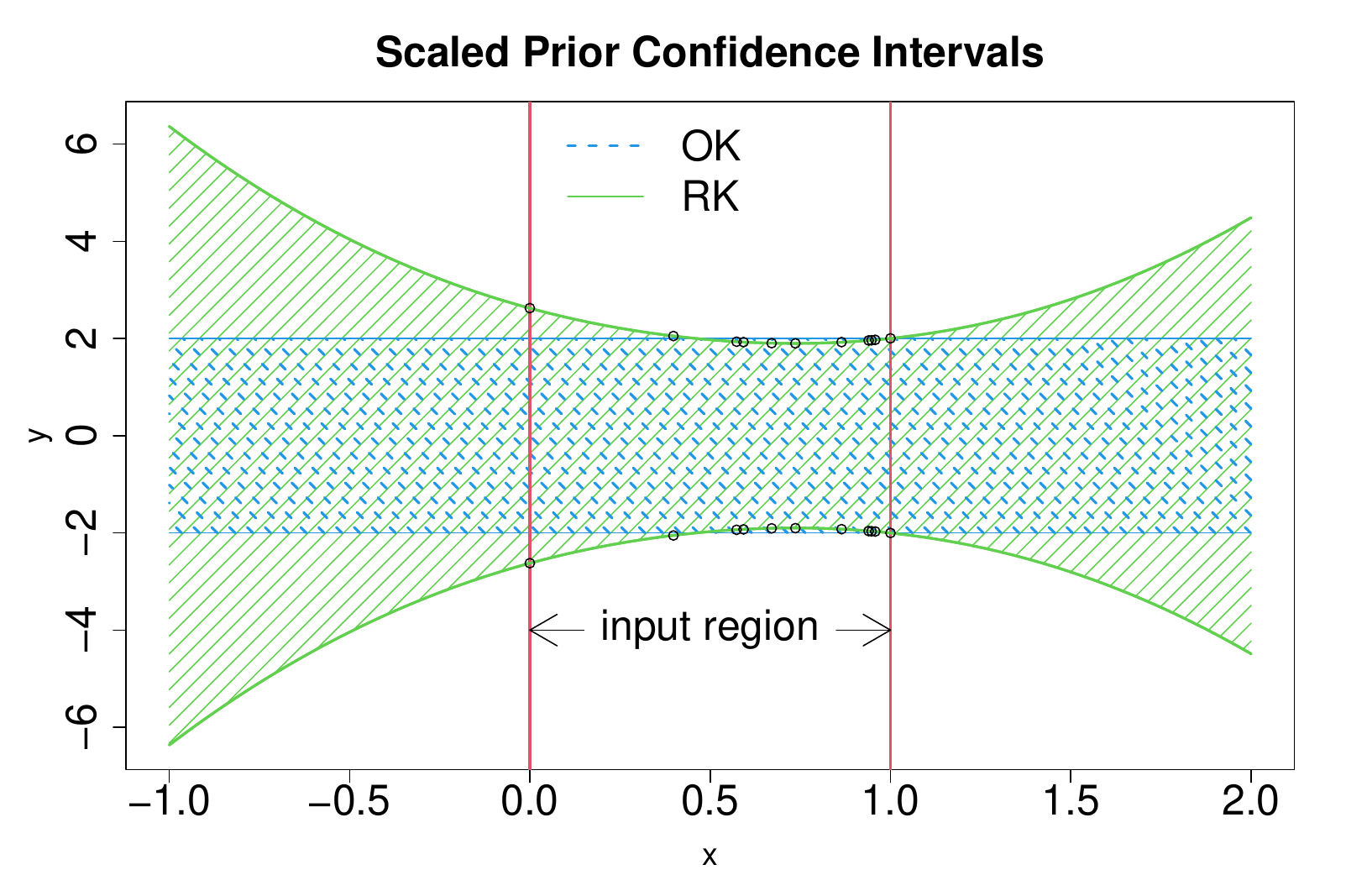} 
\caption{Scaled 95\% prior confidence regions for the function with mean centered at 0 are shown as shaded regions for OK and RK. For one of the simulations, $\pm 2/\bm r(x_i)'\hat{\bm c}$, $i=1,\ldots,11$  are plotted as points.}
\label{fig:priorci}
\end{center}
\end{figure}

For ordinary kriging, a priori, 95\% of the function is believed to lie in $[\mu-2\tau,\mu+2\tau]$, whereas for rational kriging the 95\% prior confidence interval is $[\mu-2\nu/\bm r(x)'\hat{\bm c},\mu+2\nu/\bm r(x)'\hat{\bm c}]$. They are plotted in Figure \ref{fig:priorci} by setting $\mu=0$ and $\tau=\nu=1$ for one of the simulations. We can see that they pretty much agree within the input region $[0,1]$. Outside $[0,1]$, the confidence intervals for OK remain constant, but they increase for RK. In other words, OK assigns equal ``weight'' to the whole of $\R$, whereas RK assigns more ``weight'' in the input region and less ``weight'' outside the input region. This  could be the reason why the estimates from RK are well behaved.

Additional simulations with three other one-dimensional test functions under a similar setup are reported in the Appendix A1. It can be seen from Figure \ref{fig:onedim} that in terms of prediction performance, OK and RK are comparable on the first function, OK is  better on the second function, and RK is better on the third function. On the other hand, the mean estimates from RK are much closer to the true value compared to OK for all the three functions. Some outliers are observed for RK when Gaussian correlation function is used, whereas its performance with rational quadratic is found to be much more stable.


\subsection{Universal kriging}
The universal kriging model is given by
\[y(\bm x)=\bb \beta'\bm f(\bm x)+\tau Z(\bm x),\]
where $\bm f(\bm x)=(f_0(\bm x),\ldots, f_m(\bm x))'$ is a set of known functions, $\bb \beta$ a set of unknown parameters, and $Z(\bm x)$ is a second-order stationary stochastic process with mean zero, variance $1$, and correlation function $R(\cdot)$. Ordinary kriging is a special case of universal kriging with $m=0$ and $f_0(\bm x)=1$.

The rational version of the universal kriging can be defined as
\[y(\bm x)=\bb \beta'\bm f(\bm x)+\frac{\nu}{\bm r(\bm x)'\bm c}Z(\bm x),\]
where $\bm c$ is chosen as in (\ref{eq:chat}). As before, assume $Z(\bm x)\sim GP(0,R(\cdot))$ and a noninformative prior for $\bb \beta$: $p(\bb \beta)\propto 1$. Then, the posterior distribution of the function can be obtained as \citep{Santner2003}
\begin{equation}
    y(\bm x)|\bm y\sim N\left(\hat{\bb \beta}'\bm f(\bm x),s^2(\bm x)\right),
\end{equation}
where
\begin{eqnarray*}
    \hat{\bb \beta}&=&\{\bm F' \bb\Sigma^{-1} \bm F\}^{-1}\bm F' \bb\Sigma^{-1} \bm y,\\ 
    s^2(\bm x) &=&\nu^2\left[ \frac{1-\bm r(\bm x)'\bm R^{-1}\bm r(\bm x)}{\{\bm r(\bm x)'\hat{\bm c}\}^2}+ \bm h(\bm x)'\{\bm F' \bb\Sigma^{-1}\bm F\}^{-1} \bm h(\bm x) \right],
\end{eqnarray*}
where $\bm F$ is the $n\times (m+1)$ regression model matrix, $\bb \Sigma=diag(\bm 1/\bm R\hat{\bm c})\bm R diag(\bm 1/\bm R\hat{\bm c})$, and $\bm h(\bm x)=\bm f(\bm x)-\bm F' diag(\bm R\hat{\bm c})\bm R^{-1}\bm r(\bm x)/\bm r(\bm x)'\hat{\bm c}$. The unknown parameters can be estimated using empirical Bayes:
\begin{eqnarray*}
    \hat{\nu}^2 &=& \frac{1}{n-m-1}(\bm y-\bm F\hat{\bb\beta})'\bb\Sigma^{-1} (\bm y-\bm F\hat{\bb\beta}),\\
    \hat{\bb \theta} &=&\argmin_{\bb \theta}\left\{ (n-m-1)\log\hat{\nu}^2+\log |\bm R|-2\sum_{i=1}^n \log (\bm R_i\hat{\bm c}) +\log |\bm F' \bb\Sigma^{-1}\bm F|\right\}.
\end{eqnarray*}

\begin{figure}[h]
\begin{center}
\includegraphics[width = .8\textwidth]{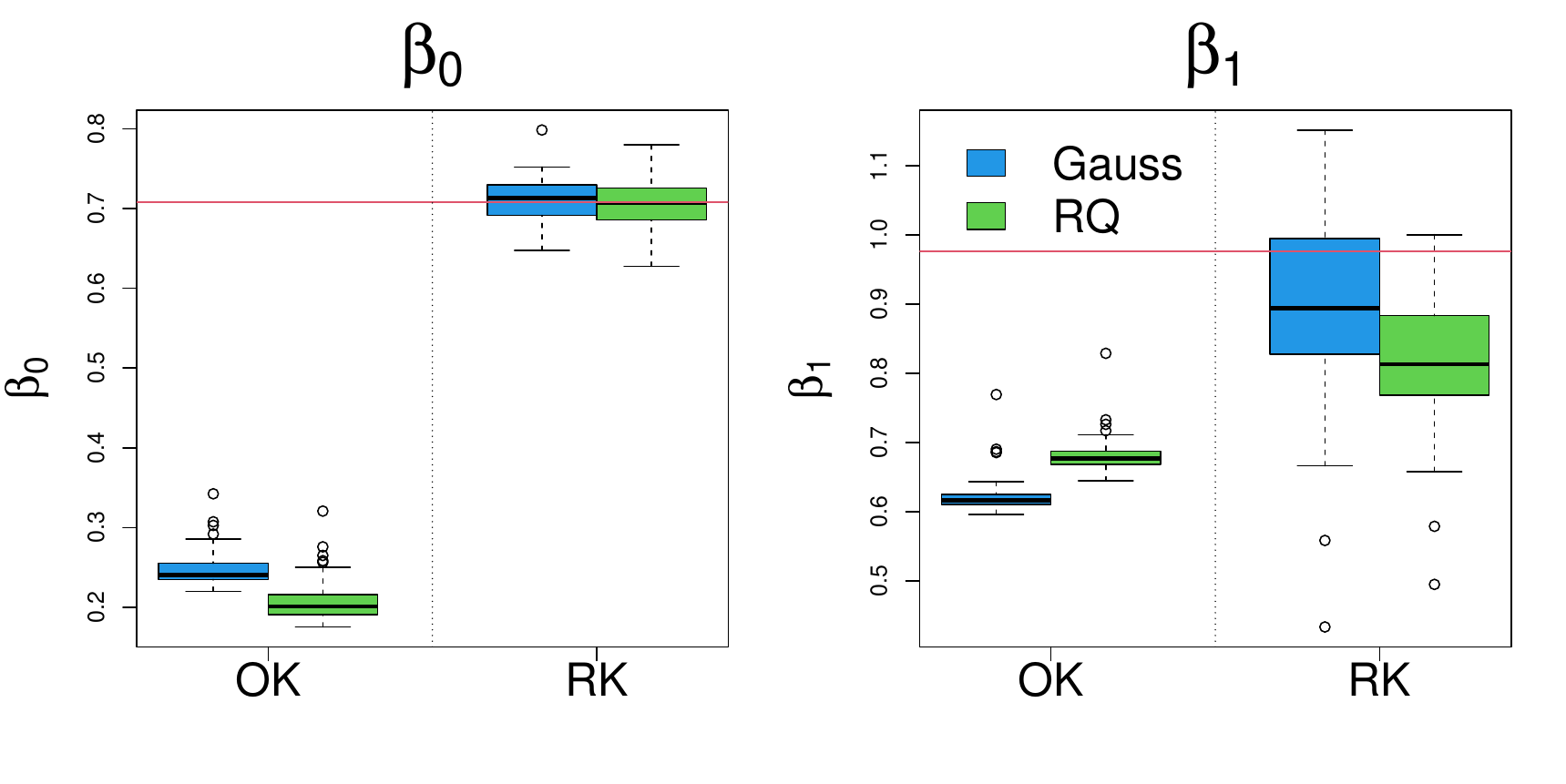} 
\caption{Results of simulation using $y=sin(2x)$ with a universal kriging model. Boxplots of $\hat{\beta}_0$ and $\hat{\beta}_1$ from the original universal kriging and the universal version of rational kriging are shown. The true least squares estimates of the two parameters are shown as red lines.}
\label{fig:uk}
\end{center}
\end{figure}

Consider a simple function $y=\sin (2x)$ for $x\in [0,1]$. The simulation in the previous section is repeated with $n=30$ using a universal kriging model having mean $E\{y(x)\}=\beta_0+\beta_1(x-.5)$. The GLS estimates of $\beta_0$ and $\beta_1$ are plotted in Figure \ref{fig:uk}. The results of RMSE and IS are omitted for brevity. The ``true'' values of the two parameters can be obtained as 
\[(\beta_0^*,\beta_1^*)=\argmin_{(\beta_0,\beta_1)}\int_0^1 \{sin(2x)-\beta_0-\beta_1(x-.5)\}^2 dx.\]
We obtain $\beta_0^*=0.708$ and $\beta_1^*=0.976$. They are plotted as red lines in Figure \ref{fig:uk}. We can see that RK gives excellent estimates of $\beta_0$ compared to OK. There is high variability for the estimates of $\beta_1$ for RK, but on the average they are still better than those from OK. Although Theorem 1 guarantees better estimation for only a constant mean function, this example shows that improving the estimation of the overall mean can indirectly improve the estimation of all the parameters in the mean  model.

\section{Applications}
In this section, we use rational kriging in two important applications of computer experiments: emulation and calibration.

\subsection{Emulation}
Borehole function \citep{morris1993bayesian} is widely used as a test function for emulation in computer experiments. It is given by
\[y=\frac{2\pi T_u(H_u-H_l)}{\ln (r/r_w)\left[1+\frac{2LT_u}{\ln (r/r_w) r_w^2 K_w}+\frac{T_u}{T_l}\right]}, \]
where the ranges of interest for the eight variables are: $r_w\in[0.05, 0.15]$, $r\in [100, 50000]$, $T_u \in [63070, 115600]$, $H_u\in[990, 1110]$, $T_l\in[63.1, 116]$, $H_l\in[700, 820]$, $L\in[1120, 1680]$, and $K_w\in[9855, 12045]$. We  scaled the variables to $[0,1]^8$ and generated $n=10\times 8=80$ points using MaxPro design \citep{joseph2015maximum}. Both ordinary kriging (using the R package \texttt{mlegp}) and rational kriging are fitted to this data using Gaussian correlation function. The root-mean squared leave-one-out cross validation error for ordinary and rational kriging are $0.654$ and $0.403$, respectively, showing that rational kriging is better for emulating the borehole function compared to ordinary kriging. Since the borehole function is a simple analytical function, we can compute the actual errors on a large testing set. Using 1,001 uniform samples from $[0,1]^8$, we obtain the root-mean squared errors as $RMSE_{OK}=0.413$ and $RMSE_{RK}=0.267$, which agrees with the results of cross validation. Similar improvements were observed for rational kriging over ordinary kriging with rational quadratic correlation function as well.

\begin{figure}[h]
\begin{center}
\includegraphics[width = 1\textwidth]{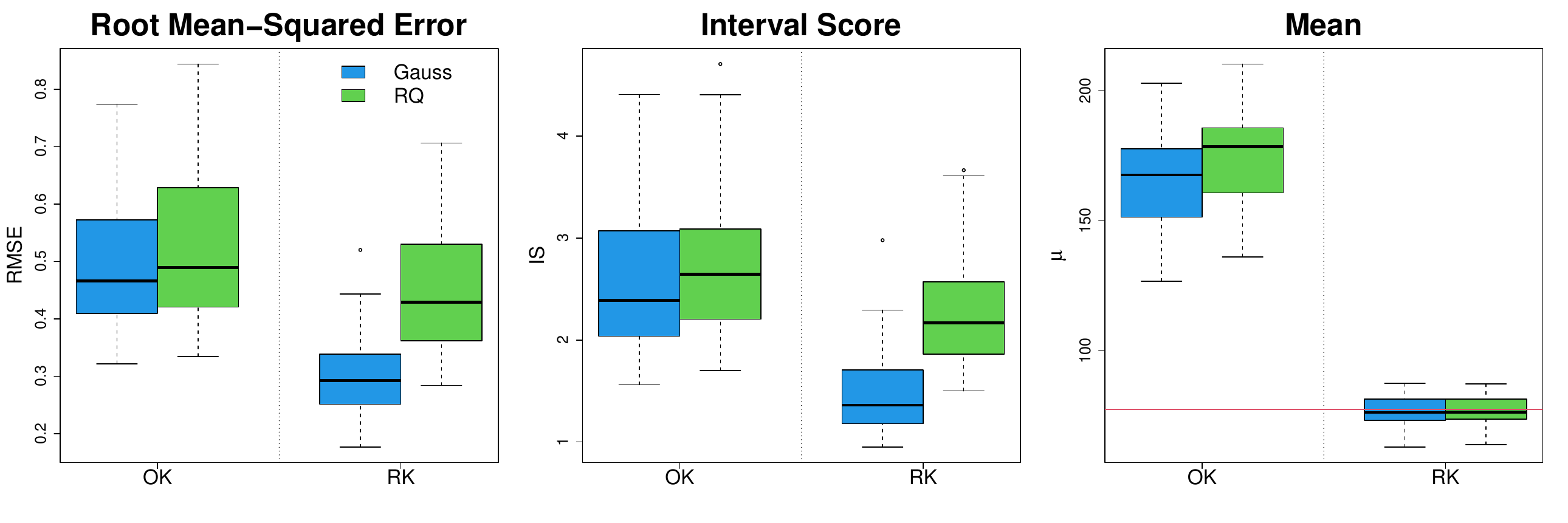} 
\caption{Boxplots of RMSE (left), IS (middle), and $\hat{\mu}$ (right) from the simulation using the borehole function. The simulation is done by randomly sampling 80 points from $[0,1]^8$. The true value $\mu^*$ is plotted as a red line in the right plot.}
\label{fig:borehole}
\end{center}
\end{figure}

We repeated the foregoing exercise 50 times  by randomly sampling 80 uniform points from $[0,1]^8$ each time and the results are summarized in Figure \ref{fig:borehole} along with interval score and the estimated mean. We can see that rational kriging outperforms ordinary kriging on both of the prediction and uncertainty quantification metrics. Using a very large uniform sample from $[0,1]^8$, the overall mean of the borehole function is estimated  to be approximately $77.74$. This is plotted as a red line in the last panel of Figure \ref{fig:borehole}. We can see that the estimates of mean from rational kriging are much closer to the true mean of the function than those from the ordinary kriging. 

Simulations using three more widely used test functions for emulation in computer experiments are reported in the Appendix. RK's prediction performance was superior to OK in all of the three cases along with better and more consistent estimates for the mean.

\subsection{Calibration}
Consider a physics-based model $y=f(\bm x;\bb \eta)$, where $\bb \eta=(\eta_1,\ldots,\eta_q)'$ are the unknown calibration parameters that need to be estimated from the real data $\{(\bm x_i,y_i)\}_{i=1}^n$. Since the physics-based model could be biased, \cite{kennedy2001bayesian} proposed to use a Gaussian process model to capture the discrepancy between the physics-based model and the data. Their model can be written as
\begin{equation}\label{eq:KOH}
    y=f(\bm x;\bb \eta)+\tau \delta(\bm x)+\epsilon,\; \delta(\bm x)\sim GP(0, R(\cdot)) \; \textrm{and}\; \epsilon\overset{iid}{\sim}N(0,\sigma^2).
\end{equation}
\cite{tuo2015efficient} have shown that this model could produce poor estimates of $\bb \eta$ because of the non-identifiability between $\bb \eta$ and $\delta(\cdot)$. Since then several proposals have appeared in the literature aimed at tackling the identifiability issue \citep{plumlee2017bayesian, gu2018scaled, tuo2019adjustments}.

Encouraged by the results of previous sections, we could consider using rational GP  in the Kennedy-O'Hagan model:
\begin{equation}\label{eq:RKOH}
    y=f(\bm x;\bb \eta)+\frac{\nu}{\bm r(\bm x)'\bm c}\delta(\bm x)+\epsilon,\; \delta(\bm x)\sim GP(0,R(\cdot)) \; \textrm{and}\; \epsilon\overset{iid}{\sim}N(0,\sigma^2).
\end{equation}
We make no claims about overcoming the identifiability issue with this new model. Our hope is that this model would produce better estimates of $\bb \eta$ than with the original Kennedy-O'Hagan model.

\begin{figure}[h]
\begin{center}
\includegraphics[width = .8\textwidth]{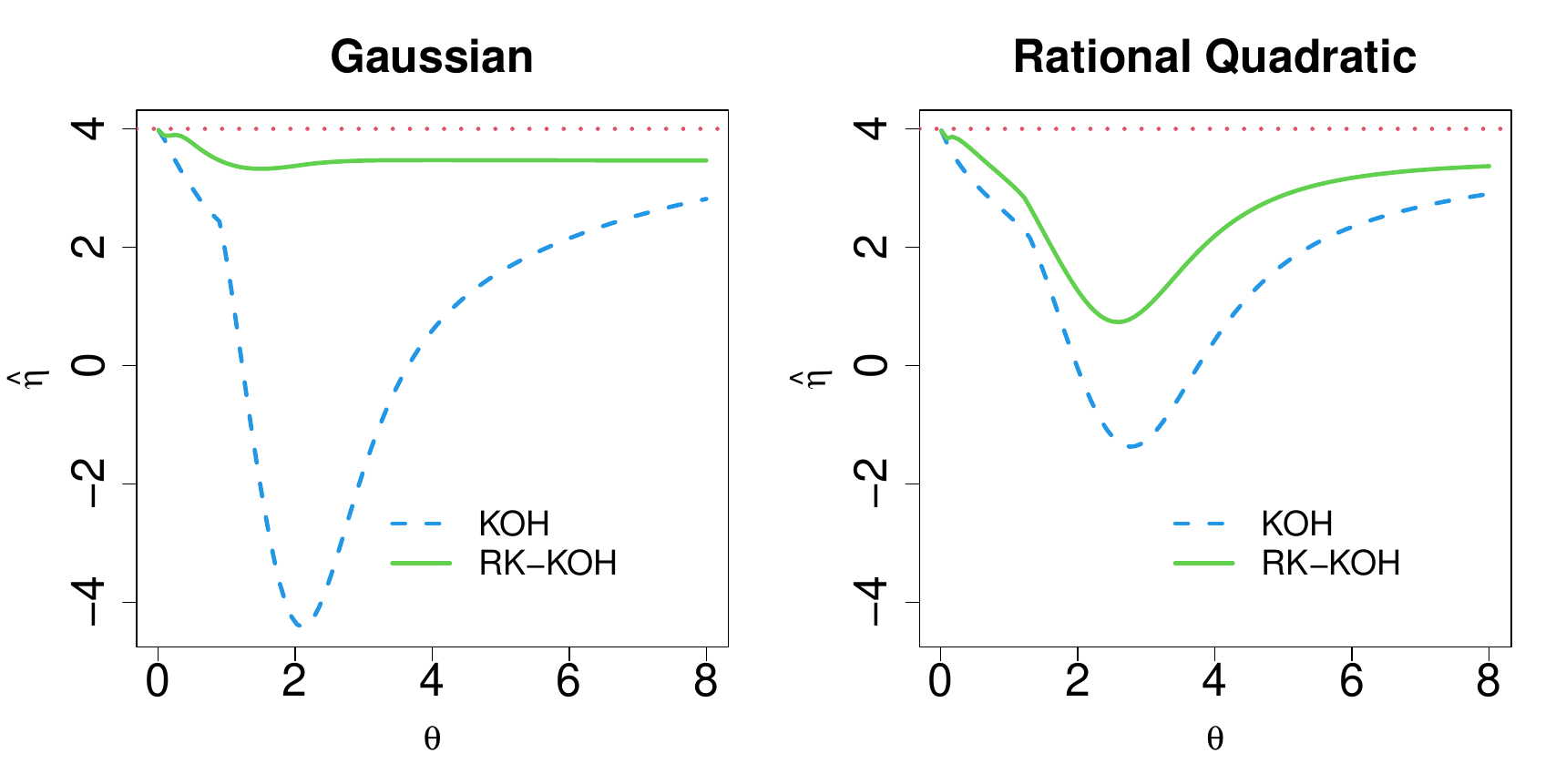} 
\caption{Plot of $\hat{\eta}$ over various values of the lengthscale parameter $\theta$ using original Kennedy-O'Hagan (KOH) model and rational version of the Kennedy-O'Hagan (RK-KOH) model. The least squares estimate of $\eta$ is shown as a red dotted line.}
\label{fig:calibration}
\end{center}
\end{figure}

Consider a simple example from \cite{plumlee2017bayesian}. Suppose $f(x;\eta)=\eta x$, but the data is generated from $y=4x+x\sin(5x)+\epsilon$ with $\epsilon \overset{iid}{\sim}N(0,0.02^2)$. Input values are generated by taking 17 equally spaced points in $[0,0.8]$. Since $f(\cdot)$ is linear in $\eta$, we can use the results of Section 3.2 with $\bb \Sigma = diag(\bm 1/\bm R\hat{\bm c})\bm R diag(\bm 1/\bm R\hat{\bm c}) + \sigma^2/\nu^2 \bm I$, where $\bm I$ is the identity matrix. Figure \ref{fig:calibration} shows the plot of $\hat{\eta}$ for various values of $\theta$ using Gaussian  and rational quadratic correlation functions. The least squares estimate of $\eta$ is around $4.0$ and is plotted in the same figure as a red dotted line. We can see that the estimates of $\eta$ from the rational version of the Kennedy-O'Hagan (RK-KOH) model are much closer to the least squares estimate than those from the original Kennedy-O'Hagan (KOH) model for both the correlation functions. Clearly there is bias from the RK-KOH, but at least the use of rational kriging seems to stabilize the parameter estimates  making $\hat{\eta}$ more robust to the misspecification of the correlation parameters. Now the ideas from \cite{plumlee2017bayesian}, \cite{gu2018scaled}, or \cite{tuo2019adjustments} could be used in conjunction with rational kriging to overcome the identifiability issue and further improve the estimates. We leave this as a topic for future research.

\section{Conclusions}
Although ordinary kriging has been widely used in statistics, the generalized least squares estimate of the mean parameter can sometimes be nonsensical. This issue has been largely ignored in the literature because prediction and uncertainty quantification can still be good if the correlation parameters are carefully tuned. Therefore, many practitioners replace the generalized least squares estimate of the mean with ordinary least squares estimate. However, this leads to inconsistencies in the modeling framework, especially when Bayesian modeling is applied. Furthermore, there are situations such as in model calibration problems, where the parameters in the mean function have physical interpretation and thus meaningful estimates of them are desired. The rational kriging proposed in this article seems to overcome these issues. It gives comparable prediction and uncertainty quantification to those of ordinary kriging, but with substantially improved estimates for the mean parameters. This is achieved by simply scaling the stochastic part of the kriging/Gaussian process by a scaling function. Therefore, the proposed method can be implemented easily in complex statistical models. Moreover, the scaling function turned out to be closely related to the first eigenfunction of the kernel used in kriging, which can be easily estimated. 

The rational kriging provides a new perspective for kriging with a nonstationary variance function. From the inception of the kriging technique, constant variance has been widely used for the stochastic component of the statistical model. This is under the assumption of stationarity that the true function is expected to lie within a constant band throughout the region of interest. This approach works well when the true function is indeed stationary. However, in practice, we never know if it is stationary or not. Thus, it makes sense to place a prior that has smaller confidence intervals in the region of data collection and that becomes bigger as the prediction point deviates from the input region of data (see Figure \ref{fig:priorci}). This introduces a fundamental shift in the way we deal with kriging and Gaussian process models.


\vspace{.25in}
\noindent {\Large\bf Acknowledgments}

\noindent This research is supported by  a U.S. National Science Foundation grant DMS-2310637.


\vspace{1in}
\noindent {\Large\bf Appendix: Additional Simulations}

\vspace{.25in}
\noindent {\large\bf A1. One-dimensional functions}
\vspace{.1in}

\begin{figure}[h!]
\begin{center}
\includegraphics[width = 1\textwidth]{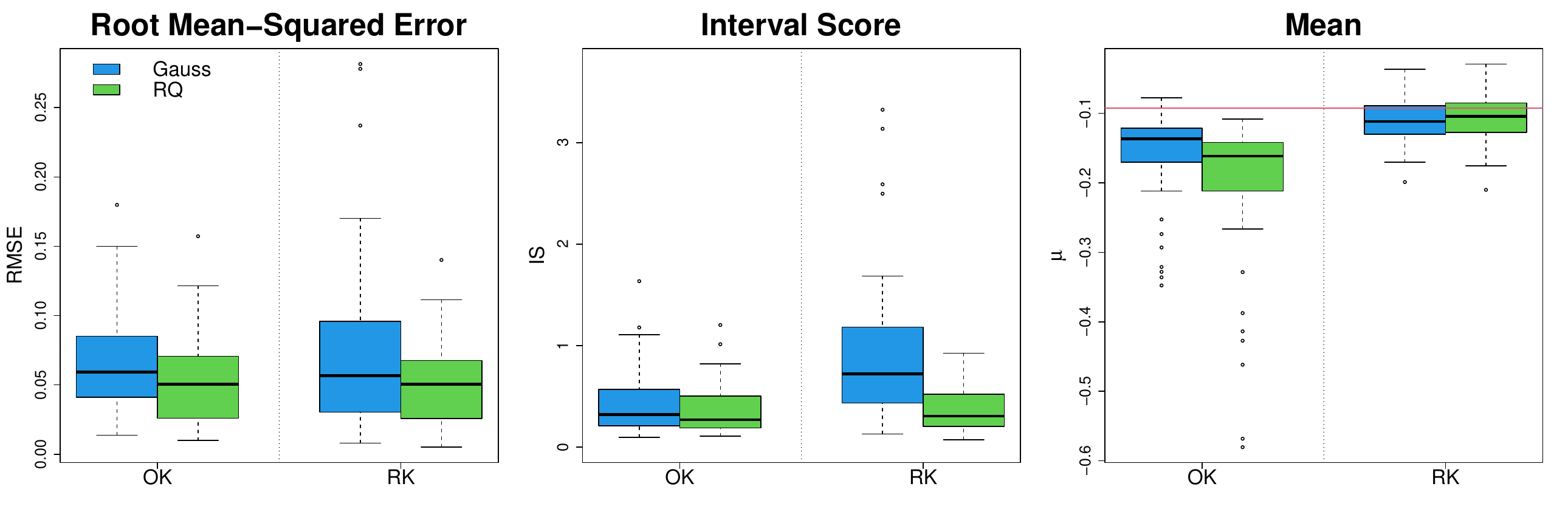} \\
\includegraphics[width = 1\textwidth]{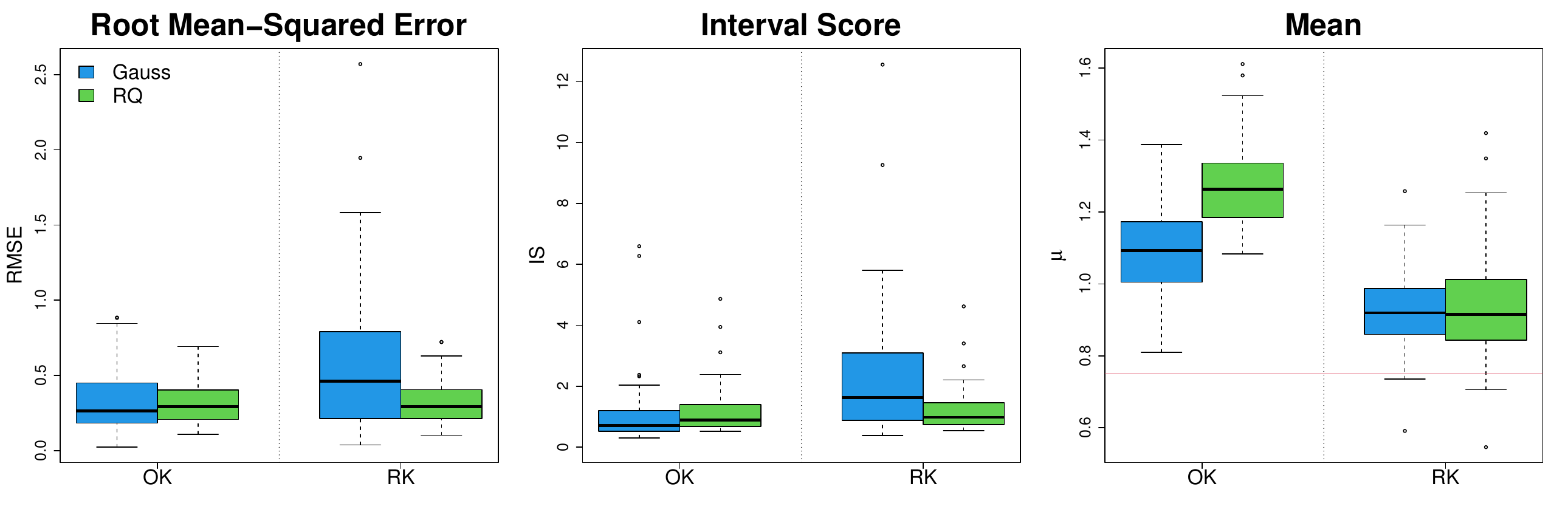}\\
\includegraphics[width = 1\textwidth]{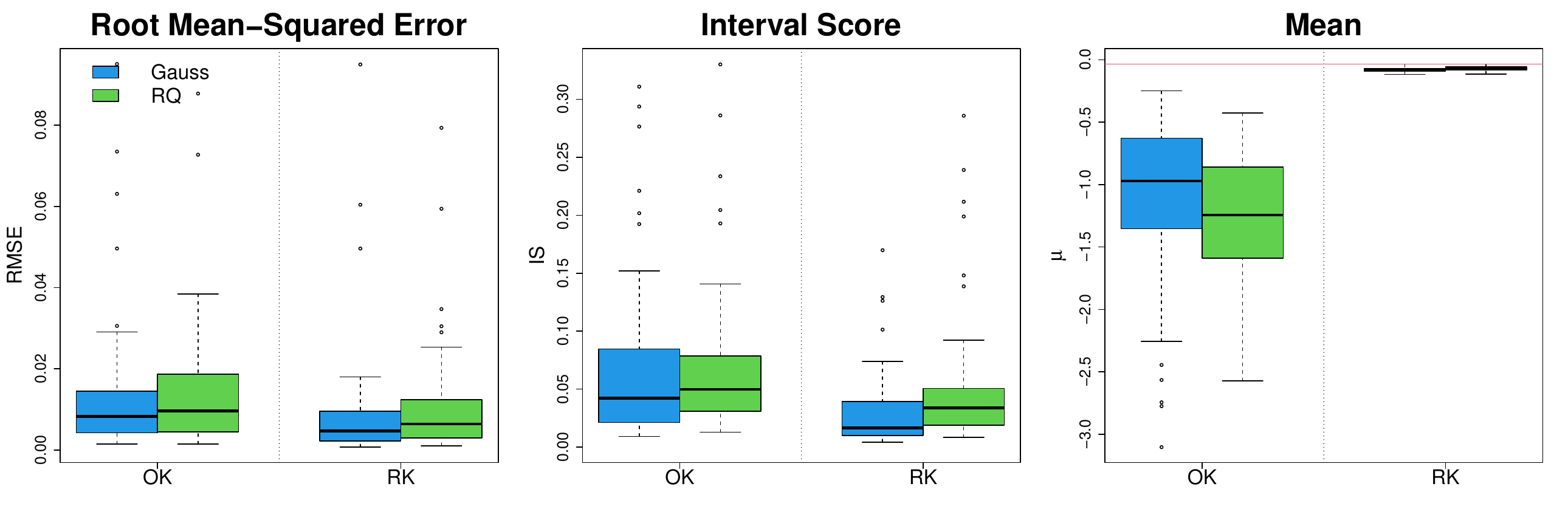}
\caption{Boxplots of RMSE (left column), IS (middle column), and $\hat{\mu}$ (right column) from the simulation using  Xiong et al function (top row), Gramacy and Lee function (middle row), and Buhmann et al function (bottom row). The simulation is done by randomly sampling $\{x_i\}_{i=1}^{n}$ from $[0,1]$. The true value $\mu^*$ is plotted as a red line in the right panels.}
\label{fig:onedim}
\end{center}
\end{figure}

The simulations in Section 3.1 with $n=30$ are repeated for three one-dimensional functions: 
\[y=\sin\{30(x-.9)^4\}\cos\{2(x-.9)\}+(x-.9)/2,\;\; x\in [0,1],\]
\[    y = \frac{\sin{10 \pi x}}{2x} + (x - 1)^4,\;\; x \in [0.5, 2.5],\]
\[ y=\frac{x^8}{\tan(1+x^2)+.5,} \;\; x \in [-1,1],\]
which are taken from \cite{xiong2007non}, \cite{gramacy2012cases}, and \cite{buhmann2020analysis}, respectively. The results are summarized in Figure \ref{fig:onedim}.

\vspace{.25in}
\noindent {\large\bf A2. Multidimensional functions}
\vspace{.1in}

\begin{figure}[h!]
\begin{center}
\includegraphics[width = 1\textwidth]{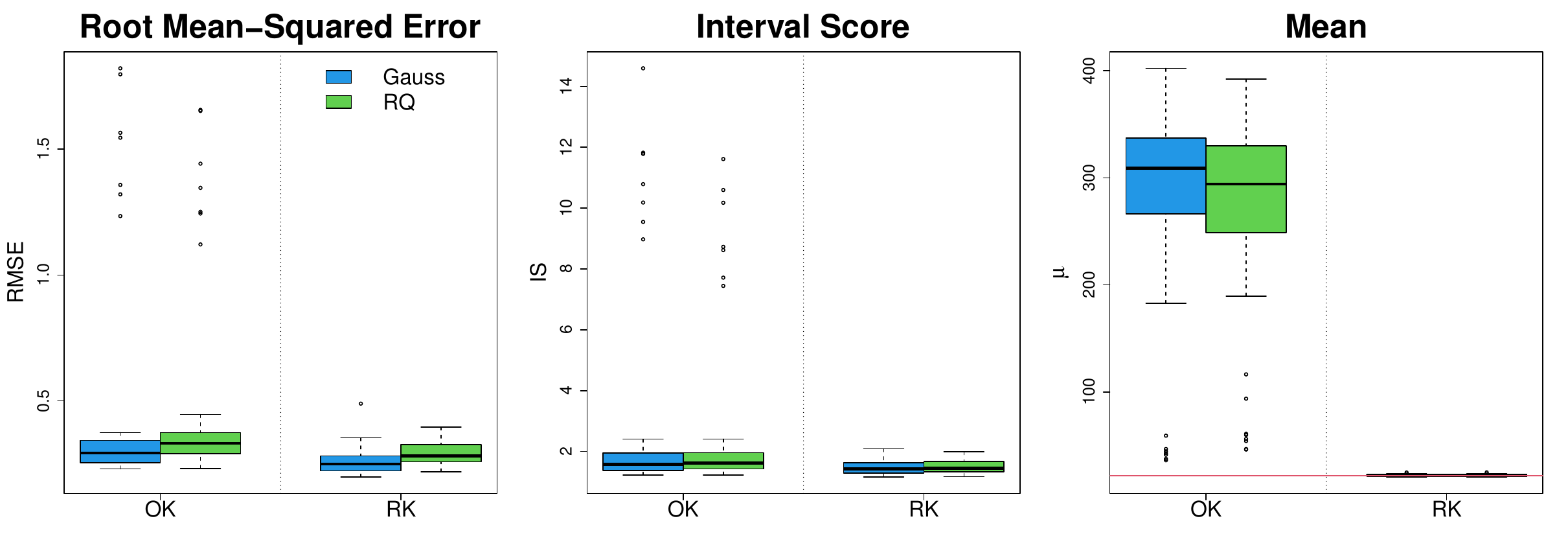} \\
\includegraphics[width = 1\textwidth]{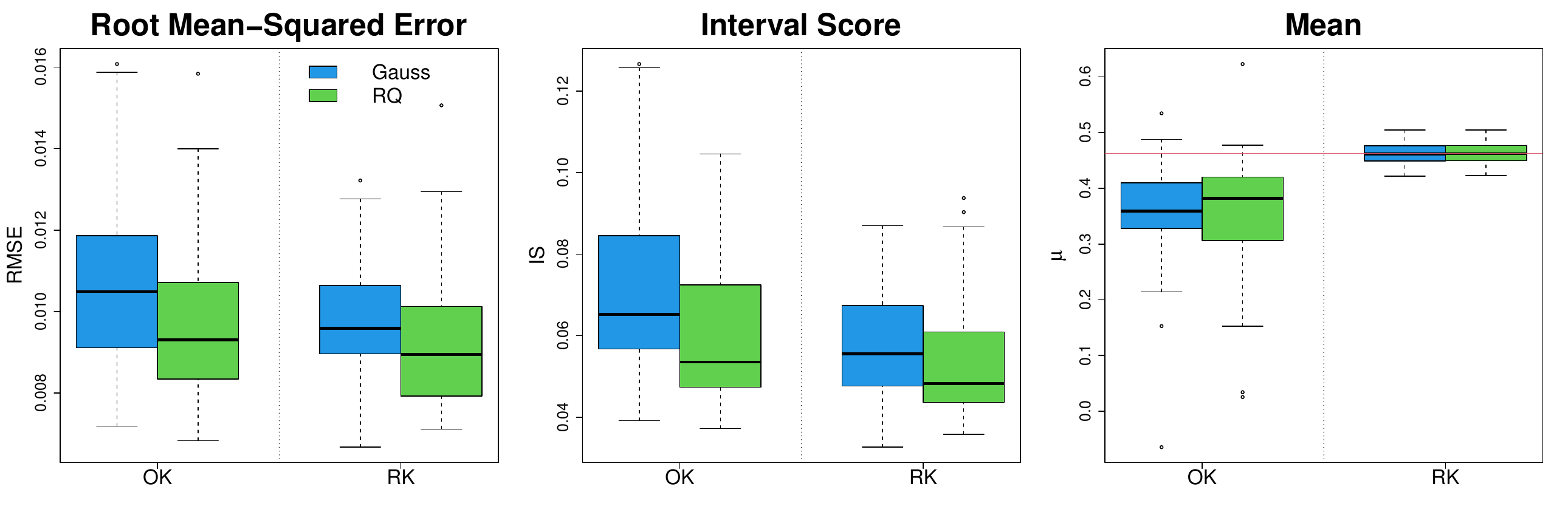}\\
\includegraphics[width = 1\textwidth]{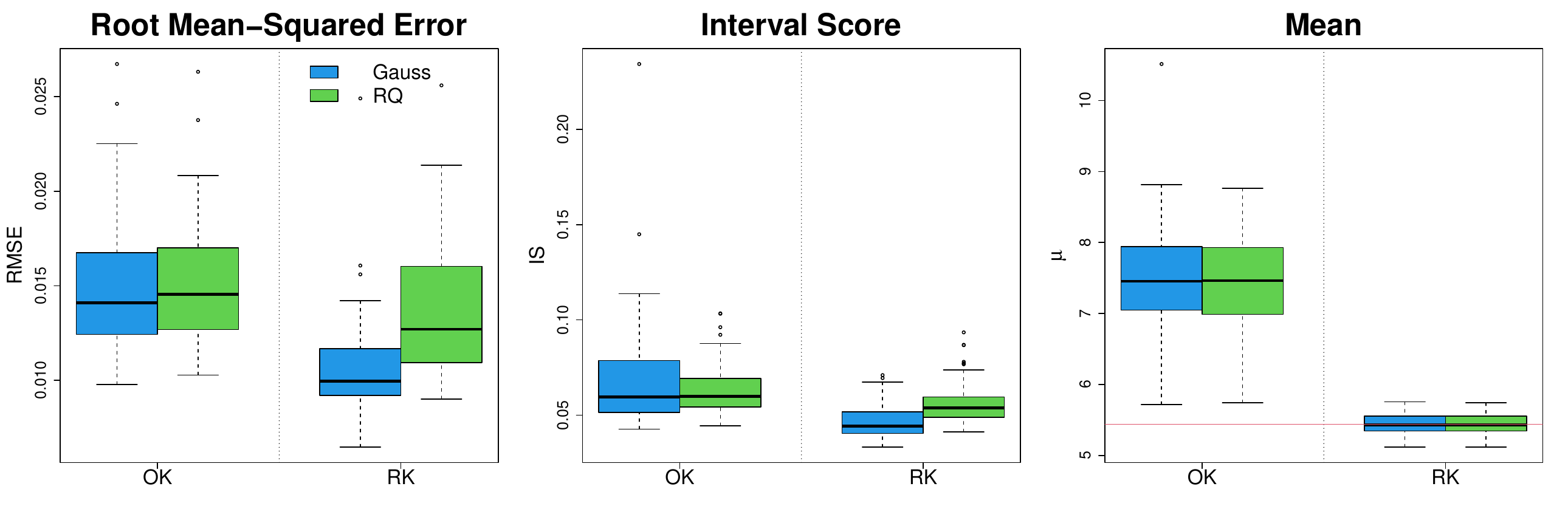}
\caption{Boxplots of RMSE (left column), IS (middle column), and $\hat{\mu}$ (right column) from the simulation using Dette-Pepelyshev function (top row), piston simulation function (middle row), and OTL circuit function (bottom row). The simulation is done by randomly sampling $n=10p$ points from $[0,1]^p$. The true value $\mu^*$ is plotted as a red line in the right panels.}
\label{fig:multidim}
\end{center}
\end{figure}

Three test functions that are widely used for emulation in computer experiments are chosen: 8-dimensional Dette-Pepelyshev function \citep{dette2010generalized}, 7-dimensional piston simulation function \citep{kenett2021modern}, and 6-dimensional OTL circuit function \citep{ben2007modeling}. The details of these functions are available at the Virtual Library of Simulation Experiments maintained by Surjanovic and Bingham https://www.sfu.ca/~ssurjano/index.html. We repeated the simulations in Section 4.1 with $n=10p$ on these three test functions. The results are summarized in Figure \ref{fig:multidim}.

\bibliography{bibliography}

\end{document}